\documentclass{elsart}
\usepackage{ifpdf}
\usepackage{graphicx,amssymb,lineno,amsmath}

\usepackage{color}                    
\newcommand{\rev}[1]{{\color{red}#1}} 

\ifpdf
\usepackage[%
  pdftitle={Instructions for use of the document class
    elsart},%
  pdfauthor={Simon Pepping},%
  pdfsubject={The preprint document class elsart},%
  pdfkeywords={instructions for use, elsart, document class},%
  pdfstartview=FitH,%
  bookmarks=true,%
  bookmarksopen=true,%
  breaklinks=true,%
  colorlinks=true,%
  linkcolor=blue,anchorcolor=blue,%
  citecolor=blue,filecolor=blue,%
  menucolor=blue,pagecolor=blue,%
  urlcolor=blue]{hyperref}
\else
\usepackage[%
  breaklinks=true,%
  colorlinks=true,%
  linkcolor=blue,anchorcolor=blue,%
  citecolor=blue,filecolor=blue,%
  menucolor=blue,pagecolor=blue,%
  urlcolor=blue]{hyperref}
\fi

\makeatletter
\def\elsartstyle{%
    \def\normalsize{\@setfontsize\normalsize\@xiipt{14.5}}
    \def\small{\@setfontsize\small\@xipt{13.6}}
    \let\footnotesize=\small
    \def\large{\@setfontsize\large\@xivpt{18}}
    \def\Large{\@setfontsize\Large\@xviipt{22}}
    \skip\@mpfootins = 18\p@ \@plus 2\p@
    \normalsize
}
\@ifundefined{square}{}{}
\makeatother

\graphicspath{{../figures/}}

\begin{document}

\begin{frontmatter}
\title{Modelling the pressure field in the vicinity
of a microphone membrane using PIV}

\author[LeMans]{O. Richoux}, \author[LeMans]{A. Degroot}, \author[LeMans]{B. Gazengel}, \author[Edinburgh]{R. MacDonald} and \author[Edinburgh]{M. Campbell}
\address[LeMans]{LAUM, CNRS, Universit\'e du Maine, Av. O. Messiaen, 72085 Le Mans, FRANCE.}
\address[Edinburgh]{Acoustics and Fluid dynamics Group, School of Physics, University of Edinburgh Kings Buildings, Edinburgh, UK}

\ead{olivier.richoux@univ-lemans.fr}

\begin{abstract}
A new approach for for estimating the acoustic pressure in the near field of a microphone based on non-intrusive direct measurement of acoustic particle velocity is proposed. 

This method enables the estimation of the acoustic pressure
inside a domain located in front of the microphone membrane. The
acoustic pressure is calculated using the acoustic particle velocity
on the frontiers of this domain and a physical model based on the
Green function of the system.\\ 

Results are obtained using the
acoustic velocity measured with Particle Image Velocimetry (PIV) in
front of a microphone excited with a plane wave inside a
rectangular waveguide. They
show that the diffraction of the plane wave by the microphone leads to an
increase of the acoustic pressure on the microphone edge in the order
of magnitude of $0.1$ dB.
\end{abstract}

\end{frontmatter}
\section{Introduction}
\label{intro}

Over the years, several methods have been developed for
microphone calibration (for a review see \cite{Zuckenwar05}). These
methods can be classified into two sorts~: (i) relative
calibration which provides an estimate of the sensitivity of a
microphone as a function of a reference sensitivity and (ii)
absolute calibration \cite{Int1} which leads to an
estimation of the sensitivity without any reference microphone.

For absolute calibration, the reciprocity technique is usually used
which provides a typical precision of around $0.05$ dB in
an enclosed field configuration. In a free field
configuration, this technique has been adapted
\cite{Barrera03,Barrera04} and standardized \cite{Int2}. Nevertheless,
free field absolute calibration suffers from numerous problems which
are not resolved at present: the location of the acoustic
center of the microphone is crucially important \cite{Wagner98}, the
generation of standing waves between the different
microphones perturbs the measurements and, generally, external 
reflections disturb the acoustic field in the vicinity of the
microphone.

Microphone calibration using non-intrusive optical techniques
  such as Laser Doppler Velocimetry is proposed by some
  authors. These studies are mainly conducted in enclosed field
  conditions, more precisely in waveguides excited with plane
  waves. Two approaches are used.

On the one hand, authors consider that the impedance of the medium is
known and measure the acoustic velocity at a single point. This
supposes that the boundary conditions of the medium are perfectly
known. We call it the "global approach".  In this case, the acoustic
pressure can be estimated at the measurement point thanks to the
impedance. Acoustic pressure can also be estimated elsewhere using a
propagation model of the system under study. Taylor \cite{Taylor81},
MacGillivray \cite{MacGillivray02,MacGillivray03} and Koukoulas
\emph{et al} \cite{Trian08} use this approach. First results are
encouraging: Taylor \cite{Taylor81} shows that this method is accurate
within $\pm 0.03$~dB at $500$~Hz, Mac Gillivray {\it et al}
\cite{MacGillivray02,MacGillivray03} reach an accuracy of around
$0.1$~dB and Koukoulas \emph{et al} \cite{Trian08} propose an accuracy
of $0.2$~dB at $170$~Hz.

On the other hand, authors consider a volume of fluid and measure the
acoustic velocity on the volume boundaries. A propagating model of the
fluid enables the estimation of the acoustic pressure everywhere
in the volume. We call this the "local approach".  For plane
waves a slice of fluid is considered and the acoustic
velocity measured at two points to estimate the acoustic pressure in
the middle of the slice.  Degroot {\it et al} \cite{Degroot08} use
this approach and show that using a $(u-u)$ probe with two LDV
measurements provides a minimum uncertainty on the pressure
estimation of $0.013$~dB for frequencies of $1360$~Hz and
$680$~Hz.

The approaches described above for enclosed field calibration could be
used in a free field. The first (global) approach only
requires the measurement of the acoustic velocity at a single point
but also requires that the impedance of the fluid for free field
conditions is known. This can only be the case if the boundary
conditions of the system are perfectly known, for example in a
semi-infinite domain which can be reproduced with a semi anecho\"ic
chamber. The second (local) approach does not require the
boundary conditions of the system under study to be known but instead
the acoustic velocity must be measured at a number of locations near
the microphone membrane, in order to estimate the acoustic pressure on
the membrane.

The local approach requires the characterization of the acoustic
nearfield of a structure by LDV or other non-intrusive
techniques. Previous work has been done by Gazengel
\textit{et al} \cite{Gazengel07}, who measure the acoustic
particle velocity in front of a loudspeaker to characterize its
acoustic radiation in a free field using the LDV
technique. Schedin \textit{et al} \cite{Schedin96} propose
measuring the acoustic particle velocity in the vicinity of a
plate using two-reference-beam double-pulsed holographic
interferometry and Moreau \textit{et al} characterize the acoustic
field in a waveguide boundary layer using LDV and PIV techniques
\cite{Moreau04}.

In this paper, we propose a preliminary study for calibrating
microphones in free field conditions using a non-intrusive
measurement technique and the local approach described
above. Section two presents the general method used for
estimating the acoustic pressure on the microphone membrane using the
acoustic velocity measured at many locations near the
membrane. Section three presents the experimental study of the acoustic
field in the close vicinity of a microphone. In this section, the
acoustic particle velocity is measured using a PIV technique. In the
fourth section, a $2$D model of the acoustic field near the microphone
in derived using the Green function of the volume under
study. Finally, the acoustic pressure field is estimated in
front of the microphone membrane using the measured acoustic velocity
and some discussion is given.

\section{General formulation of the acoustic pressure field in the vicinity of a microphone membrane}\label{model}
\begin{figure}[h!]
\centering
\includegraphics[width=10cm]{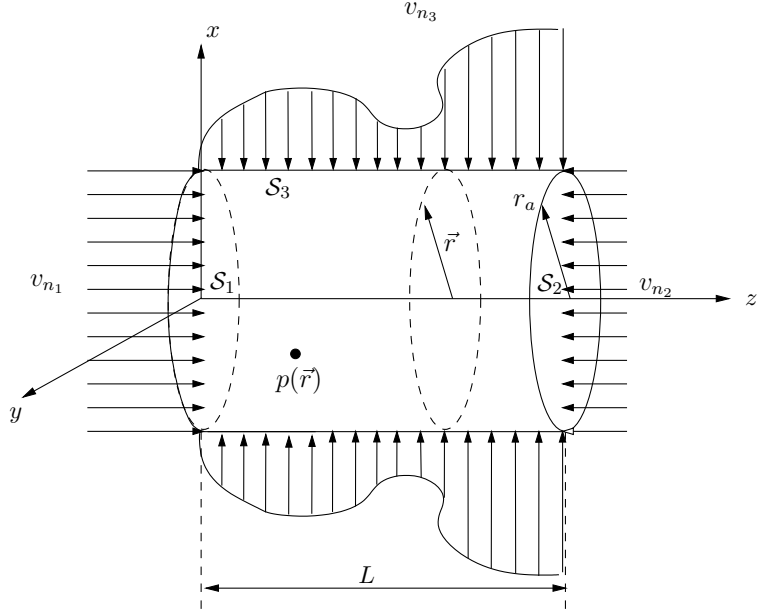}
\caption{\label{fig:studyvolume} Studied volume in front of the
microphone.}
\end{figure}

In this section, a microphone excited with a plane wave is studied. An
analytical development of the pressure field is proposed,
considering a fluid domain (air) located in front of the microphone
membrane (see Fig. \ref{fig:studyvolume}). Acoustic pressure
inside this volume can be estimated at the position
$(r,\theta,z)=(\vec{w},z)$ by means of measurement of normal velocity
on the volume boundaries with an integral formulation.

The system under study (see Fig. \ref{fig:studyvolume}) is an air
fluid column with length $L$ and a circular section of radius
$r_a$. The circular sections $(S_1,S_2)$ and the surface of the fluid
column $S_3$ are subjected respectively to the normal acoustic
velocities $v_{n_1}$, $v_{n_2}$ and $v_{n_3}$. As described in
Fig. \ref{fig:studyvolume} and according to experimental results
presented in section \ref{experimental}, the normal acoustic
velocities $v_{n_1}$ and $v_{n_2}$ are chosen to be uniform on
the circular section and $v_{n_3}$ is considered to vary on the
section $S_3$.

A general integral formulation of the acoustic pressure in the volume
$V$ is proposed hereafter.

In the frequency domain the acoustic pressure $p(\vec{r},t)$ is written as
\begin{equation}
p(\vec{r},t)=p(\vec{r}) e^{j \omega t},
\end{equation}
where $\omega$ is the acoustic pulsation defined by $\omega=2\pi
f$. In the linear acoustic approximation and in the case of a
perfect fluid, the acoustic propagation in the air fluid column $V$,
is given by
\begin{equation}
(\Delta + k2) p(\vec{r})=0,
\label{eq:propagation1}
\end{equation}
where $k=\omega/c_0$ is the wave number and $c_0$ is the sound
celerity. The boundary conditions associated with
eq. (\ref{eq:propagation1}) are written
\begin{eqnarray}
& \frac{\partial p(\vec{r})}{\partial n} = -j \omega \rho v_{n_1}(r)
\mbox{ for } r \in (0,r_a), \, \theta \in (0,2\pi), \, z=0,
\label{eq:bound_cond_p1}\\ & \frac{\partial p(\vec{r})}{\partial n} =
-j \omega \rho v_{n_2}(r) \mbox{ for } r \in (0,r_a), \, \theta \in
(0,2\pi), \, z=L, \label{eq:bound_cond_p2}\\ & \frac{\partial
p(\vec{r})}{\partial n} = -j \omega \rho v_{n_3}(r) \mbox{ for }
r=r_a, \, \theta \in (0,2\pi), \, z \in (0,L)
\label{eq:bound_cond_p3},
\end{eqnarray}
where $\rho$ is the air density and $\partial /\partial n =
\partial_n$ is the normal derivative to the surface $S$. The acoustic
pressure field $p(\vec{r})$ in the volume $V$ (described by the closed
surface $S$) is written, in this integral form, as \cite{}
\begin{equation}
p(\vec{r}) = \iiint_V G(\vec{r},\vec{r}_0) f(\vec{r}_0) dV_0 + \iint_S \left[ G(\vec{r},\vec{r}_0) \partial n_0 p(\vec{r}_0) - p(\vec{r}_0) \partial n_0  G(\vec{r},\vec{r}_0) \right] dS_0,
\label{eq:integral_form1}
\end{equation}
where the function $f(\vec{r}_0)$ describes the sources distributed
inside the volume $V$ and $G(\vec{r},\vec{r}_0)$ is the Green function
defined by the follonwing equation
\begin{equation}
(\Delta + k2) G(\vec{r},\vec{r}_0) = -\delta (\vec{r}-\vec{r}_0) \mbox{ in } V,
\label{eq:Green1}
\end{equation}
and the boundary conditions on the surface $S$. In eq. (\ref{eq:Green1}),
$\delta (\vec{r}-\vec{r}_0)$ is the Dirac distribution. The boundary
conditions for the Green function are chosen as
\begin{eqnarray}
& \partial_{n_0} G(\vec{r},\vec{r}_0) = 0 \mbox{ for } r \in (0,r_a), \, \theta \in (0,2\pi), \, z=0, \label{eq:bound_cond1}\\
& \partial_{n_0} G(\vec{r},\vec{r}_0) = 0 \mbox{ for } r \in (0,r_a), \, \theta \in (0,2\pi), \, z=L, \label{eq:bound_cond2} \\
& \partial_{n_0} G(\vec{r},\vec{r}_0) = 0 \mbox{ for } r=r_a, \, \theta \in (0,2\pi), \, z \in (0,L) \label{eq:bound_cond3},
\end{eqnarray}
to allow a description of the acoustic pressure field as a function of
the acoustic velocity on the boundaries. Using eqs. (\ref{eq:Green1}),
(\ref{eq:bound_cond1}), (\ref{eq:bound_cond2}) and
(\ref{eq:bound_cond3}), eq. (\ref{eq:integral_form1}), without any
sources ($f(\vec{r}_0)=0$), is written
\begin{equation}
p(\vec{r}) = \iint_S G(\vec{r},\vec{r}_0) \partial n_0 p(\vec{r}_0) dS_0.
\label{eq:integral_form2}
\end{equation}
Eq. (\ref{eq:integral_form2}) combined with
eqs. (\ref{eq:bound_cond_p1}), (\ref{eq:bound_cond_p2}) and
(\ref{eq:bound_cond_p3}) leads to the following expression for the
acoustic pressure in the volume $V$
\begin{equation}
p(\vec{r}) = -j \omega \rho \iint_{S} G(\vec{r},\vec{r}_0) v_{n_0} dS_0 = -j \omega \rho \sum_i \iint_{S_i} G(\vec{r},\vec{r}_0) v_{n_i} dS_i,
\label{eq:integral_form3}
\end{equation}
where $i=1, \, 2 \mbox{ and } 3$. Knowing the acoustic velocity on the
surface $S_1$, $S_2$ and $S_3$ and the Green function of the system,
the acoustic pressure can determined in the volume $V$ with the
eq. (\ref{eq:integral_form3}).

\section{Experimental characterization of the acoustic velocity field
  in the vicinity of a microphone membrane}\label{experimental}
In this section, the acoustic velocity field is estimated
experimentally using Particle Image Velocimetry (PIV) inside the
domain close to the microphone membrane as shown in figure
\ref{fig:studyvolume}. The microphone is positioned inside a
waveguide. PIV has previously been used to study acoustic flow in a
waveguide \cite{Moreau04}. The acoustic velocity field is
measured inside a plane region (laser sheet) of dimensions $2~r_a$,
$L$.

\subsection{Experimental set-up}

\subsubsection{PIV system}

The PIV system uses a pulsed Cu-Laser Oxford Lasers $LS20-50$ with a
time between pulses of 20 $\mu$s. The mean power of the
laser is $20$~W and each pulse has a duration of $25-30$ ns
with a wavelength of $510.6$~nm (green). The beam from the laser
is converted to a light sheet for delivery to the region of interest.
A CCD camera (Sensicam Double Shutter) is used for the acquisition of
the PIV images, with a resolution $1280 \times 1024$
pixels. A Berkley Nucleonics model 500 A pulse generator is used
in conjunction with National Instruments LabVIEW software to allow the
capture of PIV image pairs at any phase in the period of the acoustic
cycle. 30 measurements are made at a given phase and averaged to
estimate the velocity field. Cross-correlation and post-processing are
carried out on in-house PIV software.

\subsubsection{Acoustic system}

\begin{figure}[h!]
\centering
\includegraphics[width=12cm,draft=false]{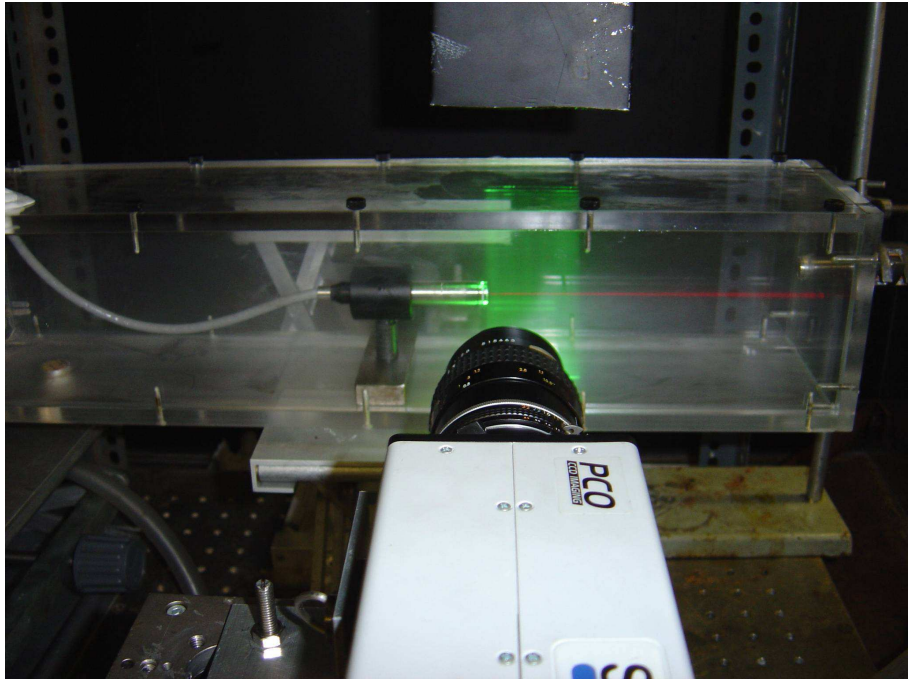}
\caption{\label{fig:exp_setup} Experimental set-up}
\end{figure}

The experimental set-up is made up of a JBL $2446H$ loudspeaker
mounted on a closed Perspex tube (with $10$ mm wall thickness)
with a length $L=0.5$~m and a square section $S = 0.1 \times
0.1$~m$^2$. The first cut-off frequency is $1720$~Hz (first transverse
mode). In the waveguide, a $1$ inch B\&K microphone is placed parallel
to the guide axis. Fig. (\ref{fig:exp_setup}) shows the acoustic
set-up and the PIV system.

\begin{figure}[h!]
\centering
\includegraphics[width=8cm]{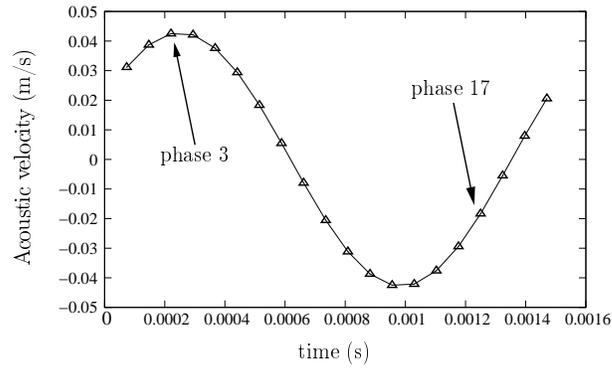}
\caption{\label{fig:phase_acoust} View of the $20$ measurement phases during one acoustic period.}
\end{figure}

A stationary plane wave with a frequency $f=680$~Hz is
established in the waveguide. The microphone membrane is located
between a node and an antinode of acoustic velocity to measure a
sufficient acoustic velocity amplitude for PIV. The light sheet
position is adjusted so as to graze the microphone membrane, in a
plane corresponding to the diameter of the membrane. The waveguide is
"seeded" using a SAFEX fog machine. The seeding is introduced during $3$ to $5$ seconds and $10$ to $15$ minutes are required before doing the measurement. The CCD Camera is
positioned perpendicular to the light sheet and focused on the
illuminated fog particles. The observation window corresponds to a
rectangular section of the volume under study (dimensions $2r_a$ x
$L$). Measurements are realized for $20$ equally spaced phase steps
in the acoustic period (see Fig. \ref{fig:phase_acoust}).

\subsection{Experimental results}\label{sec:exp_results}
\subsubsection{General analysis}
\begin{figure}[h!]
\centering
\includegraphics[width=12cm,draft=false]{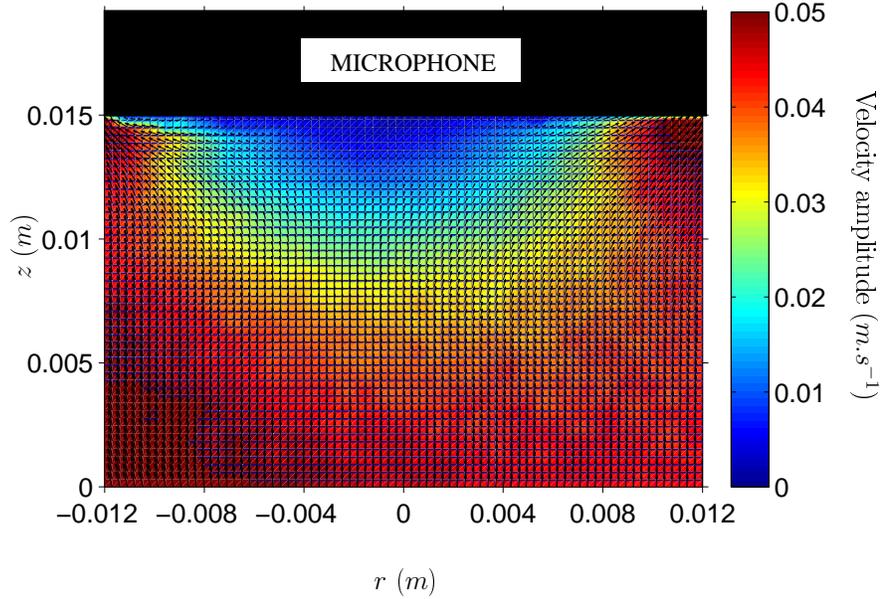}
\caption{\label{fig:velocity_phase3} View of experimental acoustic velocity field in the vicinity of the microphone for phase $3$.}
\end{figure}

The acoustic velocity field measured for the phase $3$ of the acoustic
period is shown on the Fig. \ref{fig:velocity_phase3}. The $z$-axis is
the guide axis. The dimensions of the PIV image are $0.02 \times
0.024$~m$^2$. This PIV image highlights the evolution of the acoustic
velocity field streamlines at the vicinity of the microphone
membrane. In this representation, the velocity vectors go toward the
membrane for a positive velocity amplitude. The shape of the
field lines are approximately symmetrical around the microphone axis.

\subsubsection{Longitudinal acoustic velocity}
\begin{figure}[h!]
\centering
\includegraphics[width=10cm]{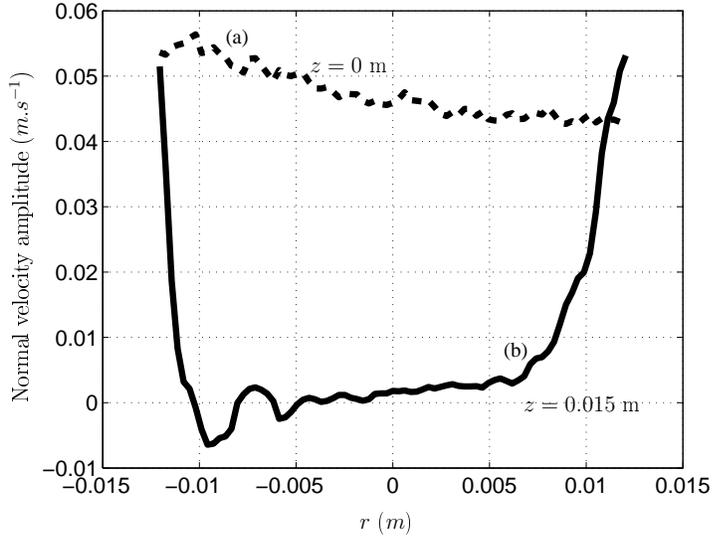}
\caption{\label{fig:amplitude_v_z} Acoustic velocity amplitude $v_z$ as a function of $r$ for $z=0$ m (a) and $z=0.015$ m (b).}
\end{figure}

Fig. \ref{fig:amplitude_v_z} shows the longitudinal acoustic velocity
$v_z$ as a function of $r$ for $z=0$~m and for $z=0.015$~m. In this figure, the microphone membrane is located at $z=0.015$~m. This
figure shows that the longitudinal velocity amplitude is almost
constant for $z=0$. The acoustic wave can therefore be considered
to be plane at this position (corresponding to surface $S_1$, see
Fig. \ref{fig:studyvolume}) in the domain under study. However, velocity $v_{n1}$ shows a variation of $\pm 5.10^{-3}$ $m.s^{-1}$ around the mean value ($5.10^{-2}$ $m.s^{-1}$), which introduces an uncertainty in the pressure estimation.

In the very closed vicinity of the microphone (the region
defined by $r \in [-0.004 ; 0.004]$~m and $z \in [0.013 ; 0.015]$~m),
the acoustic velocity field vanishes due to the membrane
stiffness. For $r=0$, the amplitude of the acoustic velocity $v_z$
decreases from $0.045$~m.s$^{-1}$ ($z=0$~m) to $2.10^{-3}$~m.s$^{-1}$
($z=0.015$~m). Assuming the membrane velocity equals the acoustic
velocity at $z=0.015$~m, this result shows that the membrane velocity
is very small compared with the acoustic velocity (ratio $\simeq
1/22$) measured at $z=0$ ($15$ $mm$ from the membrane). 
This result should be confirmed by complementary direct measurements of the membrane velocity. In further works, the microphone membrane velocity should be measured by means of a Laser Vibrometer in order to confirm this hypothesis.\\

\subsubsection{Radial acoustic velocity}
\begin{figure}[h!]
\centering
\includegraphics[width=6cm]{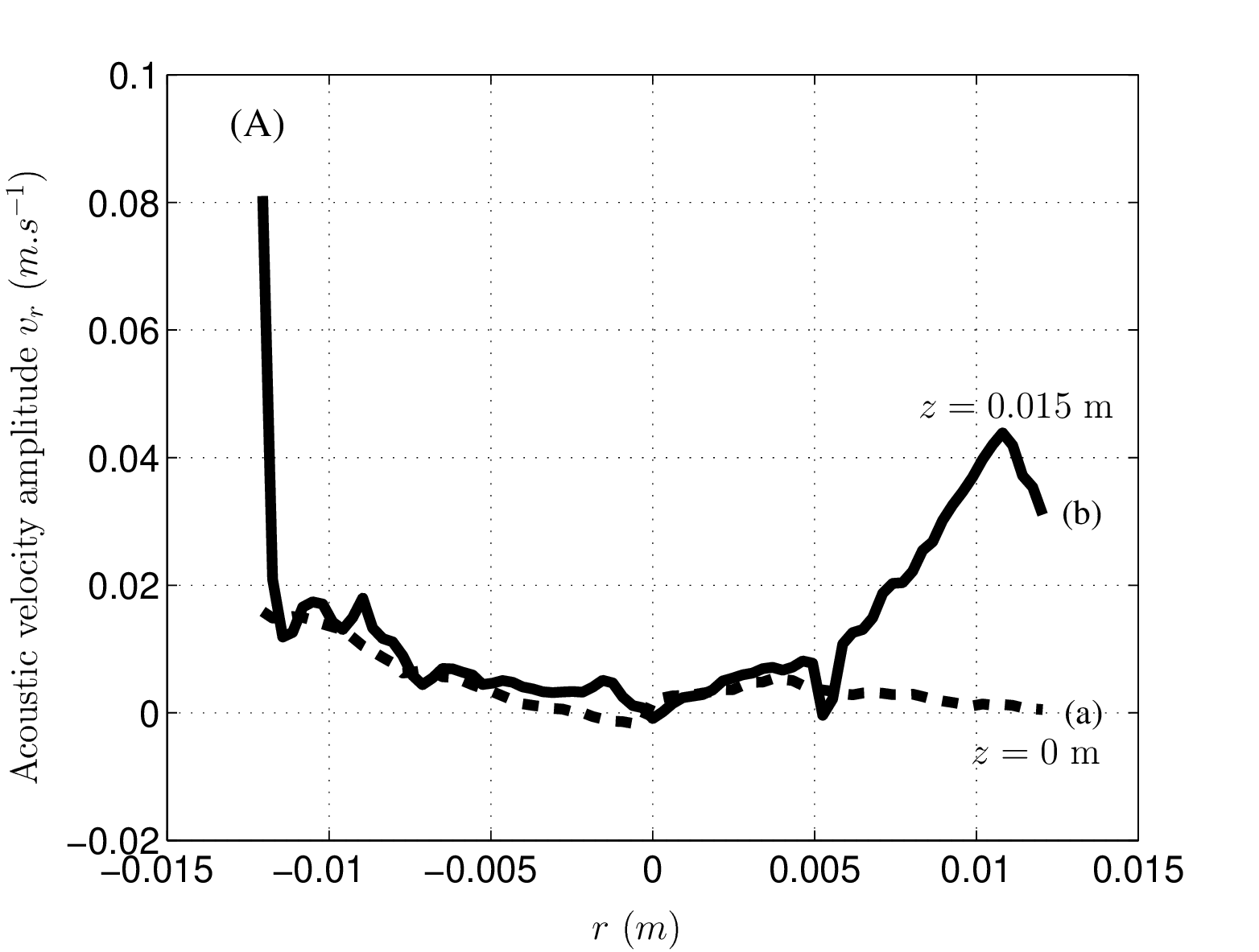}
\includegraphics[width=6cm]{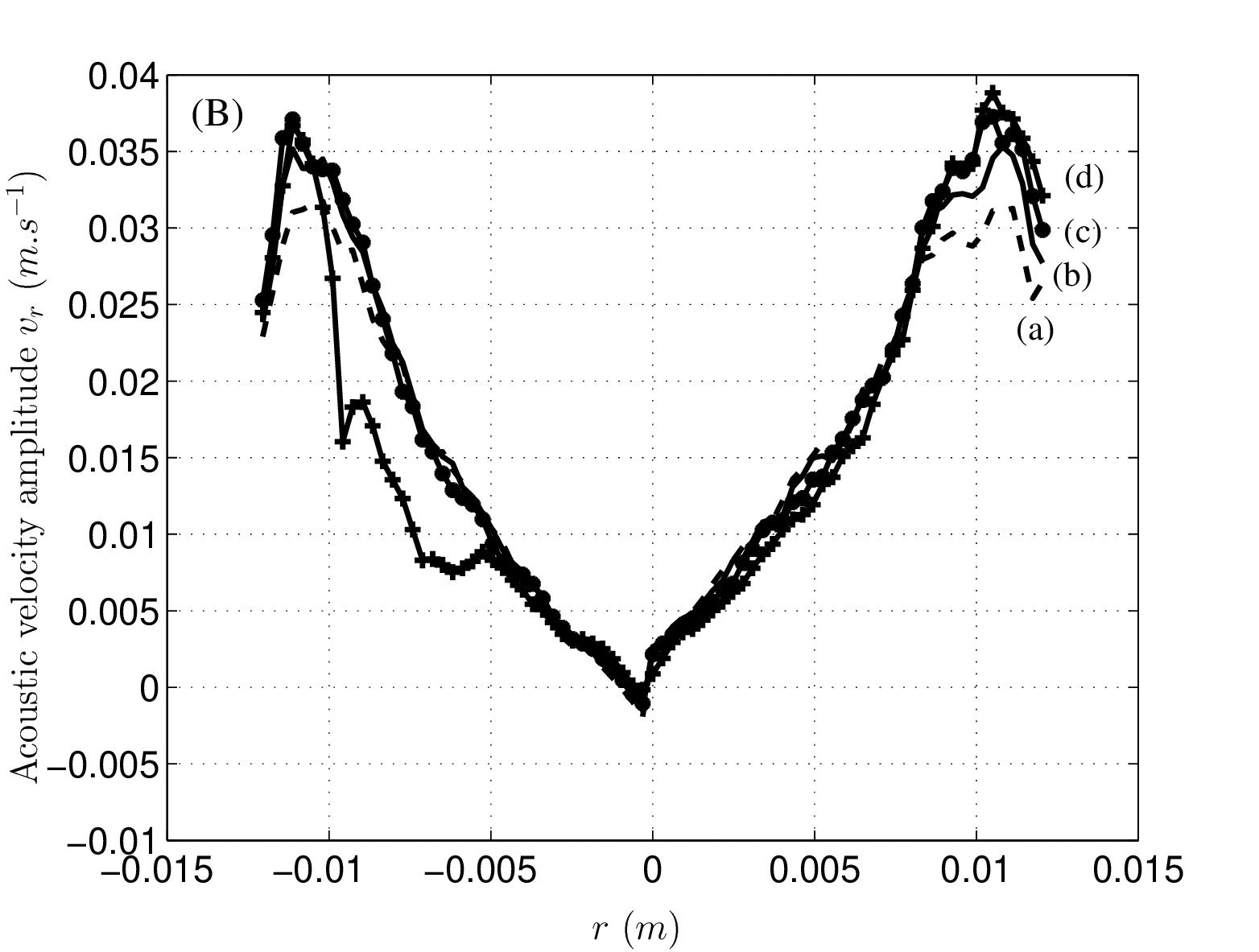}
\caption{\label{fig:vitesse_vr_fct_z_final} (A) Radial acoustic
    velocity amplitude $v_r$ as a function of $r$ for $z=0$ m (a) and
    $z=0.015$ (b). (B) Radial acoustic velocity
    amplitude $v_r$ as a function of $r$ for (a) $z=0.0136$~m ($\dots$),
    (b) $z=0.0139$~m ($--$), (c) $z=0.0142$~m ($*$) and (d) $z=0.0145$
    ($+$)}
\end{figure}

Figure \ref{fig:vitesse_vr_fct_z_final}A shows the radial acoustic
velocity amplitude $v_r$ as a function of $r$ for $z=0$~m and
$z=0.015$~m. In the very close vicinity of the membrane, the acoustic
velocity amplitude $v_r$ increases with $r$. At $z=0.015$~m and for
$r=0$~m, the radial acoustic velocity amplitude is
$v_r=1.10^{-4}$~m.s$^{-1}$ and for $r=0.011$~m,
$v_r=0.45$~m.s$^{-1}$. $v_r$ is maximum for $r=0.012$~m which
illustrates the acoustic leakage at the edge of the
microphone. In the vicinity of the
microphone, the distribution of the radial velocity amplitude is not
exactly symetric (Fig. \ref{fig:vitesse_vr_fct_z_final}A). This asymetry can be due to microphone misalignment, error in the velocity estimation with PIV or a weak air current caused by thermal effects. For greater
distance from the microphone ($z < 0.015$m), the radial velocity
amplitude becomes symetric as shown on
Fig. \ref{fig:vitesse_vr_fct_z_final}B.

\begin{figure}[h!]
\centering
\includegraphics[width=10cm]{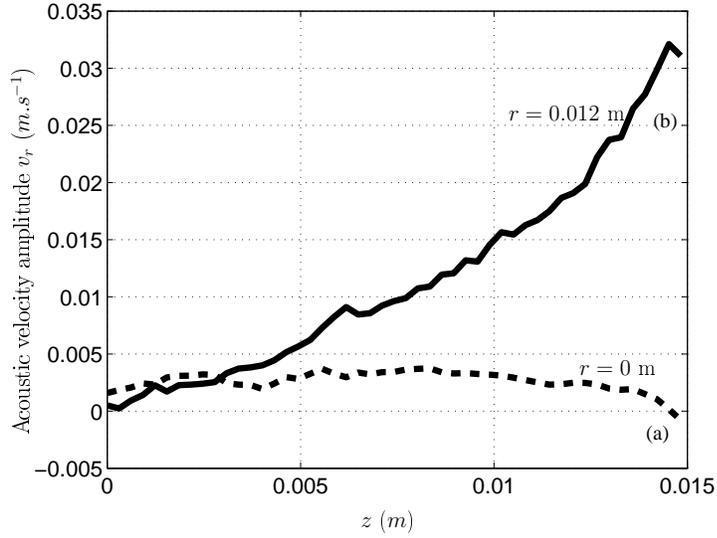}
\caption{\label{fig:vitesse_vr_final} Acoustic velocity amplitude
    $v_r$ as a function of $z$ for $r=0$ m (a) and $r=0.012$ (b).}
\end{figure}

Finally, Fig. \ref{fig:vitesse_vr_final} illustrates the radial
acoustic velocity amplitude $v_r$ as a function of $z$ for $r=0$~m and
$r=0.012$~m corresponding to surface $S_3$ shown in
Fig. \ref{fig:studyvolume}. The radial acoustic velocity amplitude
increases from zero ($r=0$~m) to $0.033$~m.s$^{-1}$ ($r=0.012$~m).
This phenomenon illustrates again the presence of acoustic leakage on the edge of the microphone and has to be taken into account in equation \ref{eq:integral_form3}.\\

Taking these experimental results into account, we consider in
the following that the longitudinal velocity $v_{n_1}$ is constant,
that the membrane velocity $v_{n_2}$ can be neglected ($v_{n_2} = 0$)
and that the radial velocity $v_{n_3}$ depends on the $z$
coordinate. The profile of $v_{n_3}$ as a function of $z$ is
estimated analytically in the following (\S \ref{results}).

\section{Analytical model of the acoustic pressure field in the vicinity of a microphone membrane}\label{model2}
In this section, we derive a specific expression of acoustic pressure
in volume $V$ using the normal velocity profiles $v_{n_1}$ and
$v_{n_3}$ estimated from experimental results (\S
\ref{experimental}). We assume an axial symmetry of the system, which
allows the derivation of a $2$D model of the acoustic field.

\subsection{Green function of the system}

The Green function $G(\vec{r},\vec{r}_0)$ solution of the problem described by eq. (\ref{eq:Green1}), can be presented as
\begin{equation}
(\Delta_{\vec{w}} + \delta_z + k2) G(\vec{w},z;\vec{w}_0,z_0)= - \delta(\vec{w}-\vec{w}_0) \delta(z-z_0) \, \mbox{ in } V,
\label{eq:Green2}
\end{equation}
where $\Delta_{\vec{w}} = (1/r)\partial_r(r\partial_r)+(1/r2)\partial_{\theta}2$. The boundary conditions (eqs. (\ref{eq:bound_cond1}), (\ref{eq:bound_cond2}) and (\ref{eq:bound_cond3})) are expressed as following
\begin{eqnarray}
& \partial_{n_0} G(\vec{r},\vec{r}_0) = \partial_{z_0} G(\vec{w},z;\vec{w}_0,z_0) = 0 \mbox{ for } \rev{z}=0, \label{eq:bound_cond_G1}\\
& \partial_{n_0} G(\vec{r},\vec{r}_0) = \partial_{z_0} G(\vec{w},z;\vec{w}_0,z_0) = 0 \mbox{ for } \rev{z}=L, \label{eq:bound_cond_G2} \\
& \partial_{n_0} G(\vec{r},\vec{r}_0) = \partial_{r_0} G(\vec{w},z;\vec{w}_0,z_0) = 0 \mbox{ for } r=r_a. \label{eq:bound_cond_G3}.
\end{eqnarray}
The Green function $G(\vec{w},z;\vec{w}_0,z_0)$ can be written as a discrete sum of eigenfunctions $\Psi_{\mu \nu} (\vec{w}) $ of the cylinder $V$ under the form
\begin{equation}
G(\vec{w},z;\vec{w}_0,z_0)= \sum_{\mu,\nu=0}^{\infty} g_{\mu \nu}(z,z_0)\Psi_{\mu \nu}(\vec{w}_0) \Psi_{\mu \nu}(\vec{w}),
\label{eq:Green3}
\end{equation}
where $g_{\mu,\nu}(z,z_0)$ are dependent on the position $z$ and $\Psi_{\mu \nu}(\vec{w})$ are solutions of the following problem
\begin{eqnarray}
& (\Delta_{\vec{w}} + k_{w \mu \nu}2)\Psi_{\mu \nu}(\vec{w}) = - \delta(\vec{w}-\vec{w}_0) \, \forall r\in (0,r_a) \mbox{ and } \forall \theta \in (0,2 \pi), \label{eq:Green4_1} \\
& \partial_{n} \Psi_{\mu \nu}(\vec{w}) =0 \mbox{ for } r=r_a \mbox{ and } \forall \theta \in (0,2 \pi).\label{eq:Green4_2}
\label{eq:Green4}
\end{eqnarray}
The eigenfunctions $\Psi_{\mu \nu}(\vec{w})=\Psi_{\mu \nu}(r,\theta)$
take the following form
\begin{equation}
\Psi_{\mu \nu}(r,\theta)= A_{\mu \nu} \cos(\mu \theta) J_{\mu}(k_{w \mu \nu} r),
\label{eq:Green5}
\end{equation}
where $J_{\mu}$ is the Bessel function and the eigenvalues $k_{w \mu \nu}$ are given by 
\begin{equation}
k_{w \mu \nu}=\frac{\gamma_{\mu \nu}}{r_a}, \mbox{ with } J'_{\mu}(\gamma_{\mu \nu})=0.
\label{eq:kwmn}
\end{equation}
The coefficients $A_{\mu \nu}$ are found using the
orthogonality of the eigenfunctions and are expressed as
\begin{equation}
A_{\mu \nu}=\frac{2}{(1+\delta_{\mu 0}) \pi r_a2 \left( 1-\frac{\mu2}{\gamma_{\mu \nu}2} \right) J2_{\mu}(\gamma_{\mu \nu})}.
\end{equation}
Using eqs. (\ref{eq:Green2}), (\ref{eq:Green3}) and
(\ref{eq:Green4_1}), the coefficients $g_{\mu \nu}(z,z_0)$ are
solutions of the following relation
\begin{equation}
(\delta_z2 + k2_{z \mu \nu}) g_{\mu \nu}(z,z_0) = \delta(z-z_0) \mbox{ for } z \in (0,L)
\end{equation}
with $k2_{z \mu \nu}=k2-k_{w \mu \nu}2$. The boundary conditions are given by
\begin{eqnarray}
& \partial_{z} g_{\mu \nu}(z,z_0) = 0 \mbox{ for } z=0, \label{eq:Green4_3} \\
& \partial_{z} g_{\mu \nu}(z,z_0) =0 \mbox{ for } z=L, \label{eq:Green4_4}
\end{eqnarray}
and the solution is written
\begin{eqnarray}
& g_{\mu \nu}(z,z_0) = -\frac{\cos(k_{z \mu \nu}z)\cos [k_{z \mu \nu} (z_0-L)]}{k_{z \mu \nu} \sin(k_{z \mu \nu}L)} \mbox{ if } z<z_0, \label{eq:Green4_5} \\
& g_{\mu \nu}(z,z_0) = -\frac{\cos(k_{z \mu \nu}z_0)\cos [k_{z \mu \nu} (z-L)]}{k_{z \mu \nu} \sin(k_{z \mu \nu}L)} \mbox{ if } z>z_0. \label{eq:Green4_6}
\label{eq:Green4}
\end{eqnarray}

\subsection{General formulation of acoustic pressure in the volume $V$}

The acoustic pressure in the volume $V$ can now be determined by means
of the Green function of the system and the normal acoustic velocity
on the surface $S$. Using eq. (\ref{eq:Green3}) in
eq. (\ref{eq:integral_form3}), the pressure is written as
\begin{equation}
p(\vec{r}) =  -j \omega \rho \sum_i \iint_{S_i} \sum_{\mu,\nu=0}^{\infty} g_{\mu \nu}(z,z_0)\Psi_{\mu \nu}(\vec{w}_0) \Psi_{\mu \nu}(\vec{w}) v_{n_i} dS_i,
\label{eq:pressure1}
\end{equation}
for $i=1, \, 2$ and $3$ corresponding respectively to the surface
$S_i$ and the normal velocities $v_{n_i}$. Setting
\begin{equation}
p_i(r,\theta,z)=-j \omega \rho \sum_{\mu,\nu=0}^{\infty} \iint_{S_i} g_{\mu \nu}(z,z_0) \Psi_{\mu \nu}(\vec{w}_0) \Psi_{\mu \nu}(\vec{w}) v_{n_i} dS_i
\label{eq:pressure2}
\end{equation}
the total pressure in the volume $V$ is written as
\begin{equation}
p(r,\theta,z)=\sum_{i=1}^{3} p_i(r,\theta,z)
\label{eq:pressure3}
\end{equation}
where $i$ indicates the considered surface of the volume. The pressure due to the different surfaces can now be calculated separately to show the influence of each surface area. 
\subsubsection{Calculation of the pressure field $p_1(r,\theta,z)$}

Using eqs. (\ref{eq:Green5}) and (\ref{eq:kwmn}) in eq. (\ref{eq:pressure2}), the pressure field $p_1(r,\theta,z)$ is expressed as
\begin{eqnarray}
p_1(r,\theta,z) = & -j \omega \rho \sum_{\mu, \nu=0}^{\infty} g_{\mu \nu}(z,0) A2_{\mu \nu} \cos(\mu \theta) J_{\mu} (\frac{\gamma_{\mu \nu}}{r_a}r) 2\pi \delta_{m0} \nonumber \\ 
& \int_{0}^{r_a} v_{n_1}(r_0) J_{\mu} (\frac{\gamma_{\mu \nu}}{r_a} r_0) r_0  dr_0 d\theta_0
\label{eq:pressure11}
\end{eqnarray}
where $v_{n_1}(r_0)$, the normal acoustic velocity field for $x=0$, is
independant of $\theta$. Taking into account the cylindrical
symmetry of the system, the acoustic pressure field $p_1$ is
also independent of $\theta$ (which implies that $\mu=0$)
and is written as
\begin{equation}
p_1(r,z)=-j \omega \rho \sum_{\nu=0}^{\infty} g_{0 \nu}(z,0) A2_{0 \nu} J_{0} (\frac{\gamma_{0 \nu}}{r_a}r) 2\pi  
\int_{0}^{r_a} v_{n_1}(r_0) J_{0} (\frac{\gamma_{0 \nu}}{r_a} r_0) r_0  dr_0 d\theta_0,
\label{eq:pressure12}
\end{equation}
with 
\begin{equation}
g_{0 \nu}(z,0)= -\frac{\cos (k_{z 0 \nu} (z-L))}{k_{z 0 \nu} \sin(k_{z 0 \nu} L)} \mbox{ for } z>0,
\label{eq:gmn}
\end{equation}
and
\begin{equation}
A2_{0 \nu}= -\frac{1}{r_a2 \pi J_0(\gamma_{0 \nu})}.
\label{eq:Amn}
\end{equation}

\subsubsection{Calculation of the pressure field $p_2(r,\theta,z)$}

The acoustic pressure field $p_2(r,\theta,z)$, considered as
independent of $\theta$ (due to the cylindrical symmetry), can be
expressed as
\begin{equation}
p_2(r,z)=-j \omega \rho \sum_{\nu=0}^{\infty} g_{0 \nu}(z,L) A2_{0 \nu} J_{0} (\frac{\gamma_{0 \nu}}{r_a}r) 2\pi  
\int_{0}^{r_a} v_{n_2}(r_0) J_{0} (\frac{\gamma_{0 \nu}}{r_a} r_0) r_0  dr_0 d\theta_0,
\label{eq:pressure21}
\end{equation}
with 
\begin{equation}
g_{0 \nu}(z,L)= -\frac{\cos (k_{z 0 \nu} z)}{k_{z 0 \nu} \sin(k_{z 0 \nu} L)} \mbox{ for } z<L,
\label{eq:gmn}
\end{equation}
and $A2_{0 \nu}$ defined by eq. (\ref{eq:Amn}).

\subsubsection{Calculation of the pressure field $p_3(r,\theta,z)$}

The normal acoustic velocity $v_{n_3}$ on the surface $S_3$,
considered as independent of $r$ and $\theta$ (due to the
cylindrical symmetry), depends only on the $z$ coordinate. The
acoustic pressure field $p_3(r,\theta,z)$, considered as
independent of $\theta$ (due to the cylindrical symmetry),
is expressed as a function of the normal acoustic velocity $v_{n_3}(z)$
on the surface $S_3$ in using eqs. (\ref{eq:Green5}) and
(\ref{eq:kwmn})
\begin{equation}
p_3(r,z)=-j \omega \rho \sum_{\nu=0}^{\infty}  A2_{0 \nu} J_{0} (\gamma_{0 \nu}) J_{0} (\frac{\gamma_{0 \nu}}{r_a}r) 2\pi r_a
\int_{0}^{L} v_{n_3}(z_0)  g_{0 \nu}(z,z_0) dz_0.
\label{eq:pressure31}
\end{equation}
The expression for $g_{0 \nu}(z,z_0)$ depends on the acoustic
velocity profile $v_{n_3}(z)$ on the surface $S_3$.

\subsection{Calculation of acoustic pressure in the volume $V$}

Using eqs. (\ref{eq:pressure3}), (\ref{eq:pressure12}), (\ref{eq:pressure21}) and (\ref{eq:pressure31}) and, as shown by the experimental results (section \ref{sec:exp_results}), considering a incident plane wave on the surface $S_1$ and $v_{n_2}(r)=0$ on the surface $S_2$, the total pressure field in the volume $V$ can be written as
\begin{eqnarray}
p(r,z)= j \omega \rho & \frac{\cos[k(z-L)]}{k \sin{kL}}v_{n_1}  + j \frac{2 \omega \rho}{r_a} \left( \sum_{\nu =0 }^{\infty} \frac{J_{0} (\frac{\gamma_{0 \nu}}{r_a}r)}{J_{0} (\gamma_{0 \nu})} \frac{1}{k_{z 0 \nu} \sin (k_{z 0 \nu} L)} \right. \nonumber  \\ 
& \left\lbrace  \cos[k_{z 0 \nu} (z-L)] \int_{0}^{z} v_{n_3}(z_0) \cos[k_{z 0 \nu} z_0]  dz_0 \right. \nonumber \\  
& \left. \left. + \cos[k_{z 0 \nu} z] \int_{z}^{L} v_{n_3}(z_0) \cos[k_{z 0 \nu} (z_0-L)]  dz_0 \right\rbrace \right),
\label{eq:pressuretotal1}
\end{eqnarray}
where $A_{0 \nu}2$ and $g_{0 \nu}(z,z_0)$ have been substituted by their expressions given by eqs. (\ref{eq:Amn}), (\ref{eq:Green4_3}) and (\ref{eq:Green4_4}).

To express the total pressure field in the volume $V$, the acoustic
velocity profile $v_{n_3}(z)$ on the surface $S_3$ must be
determined. The experimental results (section
\ref{sec:exp_results}) suggest that this profile can be modeled by a
parabolic curve defined as
\begin{equation}
v_{n_3}(z)=- \eta z2 \mbox{ for } z \in [0,L] \mbox{ and } \eta>0.
\label{eq:profile}
\end{equation}

Finally, the calculation of the integral function in eq. (\ref{eq:pressuretotal1}) using eq. (\ref{eq:profile}) leads to the following result for the total pressure in the volume $V$ (see appendix \ref{appendix2})
\begin{eqnarray}
p(r,z)= & j \omega \rho \frac{\cos[k(z-L)]}{k \sin{kL}} v_{n_1} - j \frac{2 \omega \rho}{r_a} \left( \sum_{\nu =0 }^{\infty}  \frac{J_{0} (\frac{\gamma_{0 \nu}}{r_a}r)}{J_{0} (\gamma_{0 \nu})} \frac{\eta}{k2_{z 0 \nu} \sin (k_{z 0 \nu} L)} \right. \nonumber  \\ 
& \times \left. \left\lbrace  z2 \sin (k_{z 0 \nu} L) + \frac{2}{k_{z 0 \nu}} \left[-\frac{1}{k_{z 0 \nu}} \sin(k_{z 0 \nu} L) + L \cos(k_{z 0 \nu} z)  \right] 
\right\rbrace \right),
\label{eq:pressuretotal2}
\end{eqnarray}
where $k=k_{z00}$. 

Now, the characterization of the acoustic pressure field in the
vicinity of the microphone (volume $V$) can be made with the
knowledge of the normal acoustic velocities on the volume boundaries.

\section{Results and discussion}
\label{results}

\begin{figure}[h!]
\centering
\includegraphics[width=10cm]{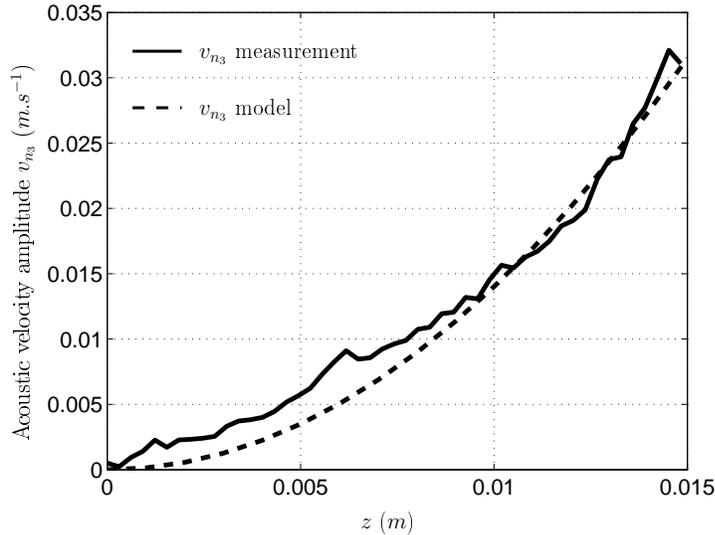}
\caption{\label{fig:comp_model_exp_v3} Normal acoustic velocity
  amplitude on the surface $S_3$ as a function of $z$,
  experiment ($-$) and model ($\cdots$) defined by
  eq. (\ref{eq:profile}).}
\end{figure}

The calculation the acoustic pressure in front of the microphone
requires the determination of the coefficient of the parabolic
curve describing the variation of the normal acoustic velocity on the
surface $S_3$. This coefficient $\eta$ defined in
eq. (\ref{eq:profile}) is determined by a minimization method.
Fig. \ref{fig:comp_model_exp_v3} shows a comparison between the
experimental data and the model (defined by eq. (\ref{eq:profile}))
with $\eta=135$~m$^{-1}$.s$^{-1}$. PIV measurements of the acoustic
velocity amplitude normal to $S_3$ have been obtained for the phase
$3$ of the acoustic period (see Fig. \ref{fig:phase_acoust}).

\begin{figure}[h!]
\centering
\includegraphics[width=10cm]{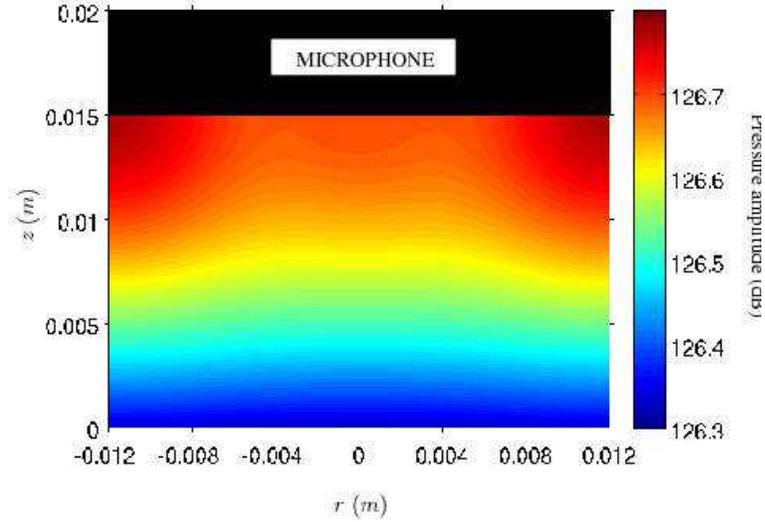}
\caption{\label{fig:pression2} Calculation of acoustic pressure amplitude in the vicinity of the microphone using eq.~(\ref{eq:pressuretotal2}) with $\nu=2$.}
\end{figure}

\begin{figure}[h!]
\centering
\includegraphics[width=10cm]{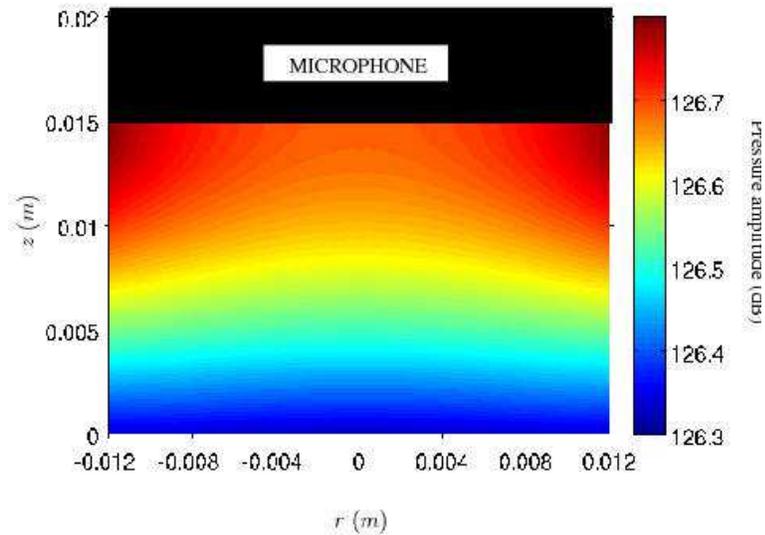}
\caption{\label{fig:pression90} Calculation of acoustic pressure amplitude in the vicinity of the microphone using eq.~(\ref{eq:pressuretotal2}) with $\nu=90$.}
\end{figure}

Figs.~\ref{fig:pression2} and \ref{fig:pression90} show the
calculation of the pressure amplitude in the volume $V$ for
$\nu=2$ modes and for $\nu=90$ modes respectively ($\nu > 0$ corresponds to radial modes defined in eq. \ref{eq:pressuretotal2}). The influence of
the number of modes on the calculation of the acoustic pressure is
clearly shown by comparing these two figures. The diffraction effects
on the acoustic wave due to the microphone are visible
even with $2$ modes but this phenomenon is described more precisely
when the modes number increases. According to the experimental
results, the presence of the microphone distorts the velocity
field streamlines allowing an acoustic leak on the border of the
microphone (Fig.~\ref{fig:velocity_phase3}). This phenomenon leads to
an increase of the acoustic pressure amplitude on the edge of the
microphone. The acoustic pressure amplitude is not constant along the
microphone membrane, varying from $126.65$ dB at the
microphone center to $126.8$ dB at the microphone border. Due to
the volume dimension and to the acoustic frequency ($f=680$~Hz), only
the plane mode is propagative and all the higher modes are
evanescent. The result of the calculation shows that the microphone
edge creates diffraction of the plane wave which leads to the
redistribution of the acoustic energy of all the modes and
consequently to the excitation of higher modes.

\section{Conclusion}
A microphone subjected to a plane wave has been studied. Using a propagation model in a fluid domain located
close to the microphone membrane and measuring the acoustic velocities
on the boundary on the domain provides an estimate of the
acoustic pressure on the microphone membrane. This model assumes that
the normal acoustic velocity on the microphone membrane is
uniformly zero and considers the incident wave as plane.

This preliminary study of the acoustic pressure field in the very
close vicinity of a microphone highlights an acoustic pressure
gradient along the microphone membrane in order of magnitude of
$0.1$~dB. This acoustic pressure difference existing between the center
and the edge of the membrane can be very problematic when
microphone calibration requires an uncertainty less than
$0.1$~dB.\\
The hypothesis used in this work (incident plane wave, parabolic
profile of radial velocity and motionless membrane) tend to minimize this pressure gradient
value. A numerical calculation of integrals (Green fomulation) would enable to ignore 
these hypothesis and could lead to a more realistic pressure estimation.

This preliminary study opens new horizons in microphone
calibration research, especially in free field conditions. These
first results on the acoustic pressure field in the very close
vicinity set an important question : what is measured by the
microphone since the acoustic pressure is not uniform along the
membrane ?

In the future, these results should be validated by measuring the
acoustic pressure field in the very close vicinity of the microphone
using, for instance, a small probe to not perturb the acoustic
field. The proposed model can be improved by taking into account the
membrane motion which could be estimated using laser probe
measurements.




\end{document}